\documentclass[conference]{IEEEtran}
\IEEEoverridecommandlockouts
\usepackage{cite}
\usepackage{amsmath,amssymb,amsfonts}
\usepackage{algorithmic}
\usepackage{graphicx}
\usepackage{textcomp}
\usepackage{xcolor}
\usepackage{todonotes}
\usepackage{booktabs}
\usepackage{url}
\def\BibTeX{{\rm B\kern-.05em{\sc i\kern-.025em b}\kern-.08em
    T\kern-.1667em\lower.7ex\hbox{E}\kern-.125emX}}
\usepackage{makecell}
\usepackage{multirow}
\usepackage{framed}
\usepackage{hyperref}
\usepackage{tikz}
\usetikzlibrary{calc,positioning,shapes}
\usetikzlibrary{arrows}
\usetikzlibrary{backgrounds}

\definecolor{myBlue}{rgb}{0.0,0.355,0.652}
\definecolor{myBlueBackground}{rgb}{0.6, 0.81, 0.93}
\definecolor{myGreen}{rgb}{0.390,0.695,0.285}
\definecolor{myRed}{rgb}{0.7,0,0}
\definecolor{darkcyan}{rgb}{0.0, 0.55, 0.55}

\usepackage{pifont}

\newcommand{\etal}{\textit{et~al}.~}
\newcommand{\ie}{\textit{i}.\textit{e}.,~}
\newcommand{\eg}{\textit{e}.\textit{g}.,~}
\newcommand{\wrt}{\textit{w}.\textit{r}.\textit{t}.~}
\newcommand{\cf}{\textit{cf}.~}

\newcommand{\gptnew}{\mbox{GPT-4o mini}}
\newcommand{\gptold}{\mbox{GPT-3.5 Turbo}}
\newcommand{\msphi}{\mbox{Phi-4}}

\newcommand{\copilot}{GitHub Copilot}

\newcommand{\circledone}{\ding{192}}
\newcommand{\circledtwo}{\ding{193}}
\newcommand{\circledthree}{\ding{194}}

\begin{document}

\title{Feedback Loops and Code Perturbations \\ in LLM-based Software Engineering: \\ A Case Study on a C-to-Rust Translation System\thanks{© 2026 IEEE.  Personal use of this material is permitted.  Permission from IEEE must be obtained for all other uses, in any current or future media, including reprinting/republishing this material for advertising or promotional purposes, creating new collective works, for resale or redistribution to servers or lists, or reuse of any copyrighted component of this work in other works. DOI 10.1109/SANER67736.2026.00047}}

\author{\IEEEauthorblockN{Martin Weiss}
\IEEEauthorblockA{
\textit{Otto von Guericke University}\\
Magdeburg, Germany}
\and
\IEEEauthorblockN{Jesko Hecking-Harbusch}
\IEEEauthorblockA{
\textit{Bosch Research}\\
Renningen, Germany}
\and
\IEEEauthorblockN{Jochen Quante}
\IEEEauthorblockA{
\textit{Bosch Research}\\
Renningen, Germany}
\and
\IEEEauthorblockN{Matthias Woehrle}
\IEEEauthorblockA{
\textit{Bosch Research}\\
Renningen, Germany}
}

\maketitle

\begin{abstract}
The advent of strong generative AI has a considerable impact on various software engineering tasks such as code repair, test generation, or language translation. While tools like GitHub Copilot are already in widespread use in interactive settings, automated approaches require a higher level of reliability before being usable in industrial practice. In this paper, we focus on three aspects that directly influence the quality of the results: a) the effect of automated feedback loops, b) the choice of Large Language Model (LLM), and c) the influence of behavior-preserving code changes.

We study the effect of these three variables on an automated C-to-Rust translation system. Code translation from C to Rust is an attractive use case in industry due to Rust's safety guarantees.
The translation system is based on a generate-and-check pattern, in which Rust code generated by the LLM is automatically checked for compilability and behavioral equivalence with the original C~code.
For negative checking results, the LLM is re-prompted in a feedback loop to repair its output.
These checks also allow us to evaluate and compare the respective success rates of the translation system when varying the three variables.

Our results show that without feedback loops LLM selection has a large effect on translation success.
However, when the translation system uses feedback loops the differences across models diminish. 
We observe this for the average performance of the system as well as its robustness under code perturbations. 
Finally, we also identify that diversity provided by code perturbations can even result in improved system performance.
\end{abstract}


\section{Introduction}

The advent of strong generative AI has profoundly impacted various software engineering tasks~\cite{DBLP:journals/tosem/HouZLYWLLLGW24, fan_large_2023}, including code repair~\cite{rondon2025evaluatingagentbasedprogramrepair}, test generation~\cite{alshahwan2024automated}, and language translation~\cite{DBLP:journals/corr/abs-2405-11514,DBLP:journals/corr/abs-2404-18852,zhang2024scalable}. 
Especially for code related tasks, text-based generative~AI in the form of \textit{Large Language Models (LLMs)} supports developers everyday across a wide range of code editors and coding assistants.
This impressive adoption rate from developers as well as studies on development speed~\cite{paradis2024much} show a positive impact of such assistive technologies. 
Nevertheless, there may also be a negative impact, \eg on quality~\cite{paradis2024much} and \wrt security~\cite{perry2023users}. However, for particular tasks, tools may leverage \emph{assured} LLM-based software engineering~\cite{alshahwan2024assured} where the properties of the generated target program can be automatically checked.
Software engineering tasks that are amenable to such a \textit{generate-and-check} pattern and allow developers to formulate automated checks are particularly interesting for LLM-based automation. 
Particularly, \textit{generate-and-check} enables automated \textit{feedback loops}, \ie to spend additional LLM queries to improve system performance at additional cost. 
In this work, we study various influence factors that impact this trade-off between performance and cost.
Furthermore, we study the impact of behavior-preserving code changes, so-called \textit{code perturbations}, on quality.

We evaluate these factors on the task of LLM-based translation from C to Rust,
which has become a prominent research focus~\cite{DBLP:journals/corr/abs-2405-11514,DBLP:journals/corr/abs-2404-18852,DBLP:journals/corr/abs-2409-10506,DBLP:journals/ese/HongR25,DBLP:journals/ase/ZhangZX26,darpa_tractor}.
This is because Rust eliminates some of the most critical classes of runtime errors common in~C by design but is still suitable for systems programming, making it particularly appealing for safety- and security-relevant applications.
However, manually rewriting the vast amount of existing C~code is impractical, so an automated approach is highly desirable. Traditional rule-based approaches like c2rust\footnote{\url{https://github.com/immunant/c2rust}} have the drawback that most code is put into ``unsafe'' blocks, which makes many of Rust's safety mechanisms ineffective and produces code that is hard to maintain~\cite{DBLP:journals/cacm/HongR25}. 

\begin{figure*}[!ht]
	\centering
	\resizebox{1\textwidth}{!}{\begin{tikzpicture}[		
    -stealth,
    very thick,
    >=stealth',
    node distance=5mm and 15mm,
    elem/.style={
        rectangle,
        draw=black, 
        very thick,
        text centered,
        minimum height=10mm,
        minimum width=20mm,
        rounded corners,
        align=center
    }
]
\tikzstyle{myarrows}=[line width=1mm,-triangle 45,postaction={line width=2mm, shorten >=2mm, -}]
\node[elem] (orig) {Original C code};
\node[elem, right=of orig, xshift=-7.5mm, dashed] (pert) {Perturbations \circledthree\\(optional)};
\node[elem, right=of pert, xshift=-7.5mm] (code) {C code};
\node[elem, right=of code, xshift=-7.5mm] (llm) {LLM \circledtwo};
\node[elem, below=of llm] (prompt) {Translation\\prompt};
\node[elem, right=of llm, xshift=1mm] (agent) {Feedback\\loop};
\node[elem, above right=of agent, xshift=10mm, yshift=-7.5mm] (rust) {Valid Rust\\translation};
\node[elem, below right=of agent, xshift=10mm, yshift=7.5mm] (fail) {Fail};
\node[elem, below =of agent, minimum width=16mm, xshift=-17mm] (rustc) {Rust\\compiler};
\node[elem, below =of agent, minimum width=16mm] (clippy) {Clippy\\(linter)};
\node[elem, below =of agent, minimum width=16mm, xshift=17mm] (lib) {clang\\libfuzzer};

\draw[very thick] (orig) -- (pert);
\draw[very thick] (pert) -- (code);
\draw[very thick] (code) -- (llm);
\draw[very thick] (prompt) -- (llm);
\draw[very thick] ([yshift=2mm]llm.east) -- node[above] {Rust} ([yshift=2mm]agent.west);
\draw[very thick] ([yshift=-2mm]agent.west) -- node[below] {Error(s)} ([yshift=-2mm]llm.east);
\node[elem, right=of agent, opacity=0] (hidden) {};
\node[opacity=0] (right) at ($ (agent) !.9! (hidden) $) {};

\draw[-, very thick] (agent.east) -- (right.center);
\draw[very thick] (right.center) |- (rust.west);
\draw[very thick] (right.center) |- (fail.west);
\draw[very thick] (clippy.north) -- (agent.south);
\node[opacity=0] (below) at ($ (clippy) !.45! (agent) $) {};
\draw[-, very thick] (rustc.north) |- (below.center);
\draw[-, very thick] (lib.north) |- (below.center);

\begin{pgfonlayer}{background}
\draw [-, rectangle, rounded corners, very thick]
([xshift=-4mm, yshift=6.5mm]llm.north west) 
-- node[below left,xshift=5mm] {C-to-Rust translation system} ([xshift=16.5mm, yshift=6.5mm]agent.north east)
-- ([xshift=16.5mm, yshift=-16.5mm]agent.south east)
-- ([xshift=-4mm, yshift=-16.5mm]llm.south west)
-- cycle;
\end{pgfonlayer}

\draw[-stealth, very thick, fill=white] 
    ([xshift=20mm, yshift=6.5mm]agent.north east)
    arc[start angle=0, end angle=360, radius=3.5mm];
\node at ([xshift=16.5mm, yshift=6.5mm]agent.north east) {$k$ \circledone};

\draw[-stealth, very thick, fill=white] 
    ([xshift=3.5mm]agent.north east)
    arc[start angle=0, end angle=360, radius=3.5mm];
\node at (agent.north east) {$i$ \circledone};
\end{tikzpicture}}
	\caption{Overview of our C-to-Rust translation system and the three major experiments we perform in this paper: \circledone\ impact of internal feedback loops (symbolized by $i$) and multiple runs (symbolized by $k$) on performance, \circledtwo\ effect of LLM selection on performance, and \circledthree\ robustness under code perturbations.}
	\label{fig:overview}
\end{figure*}
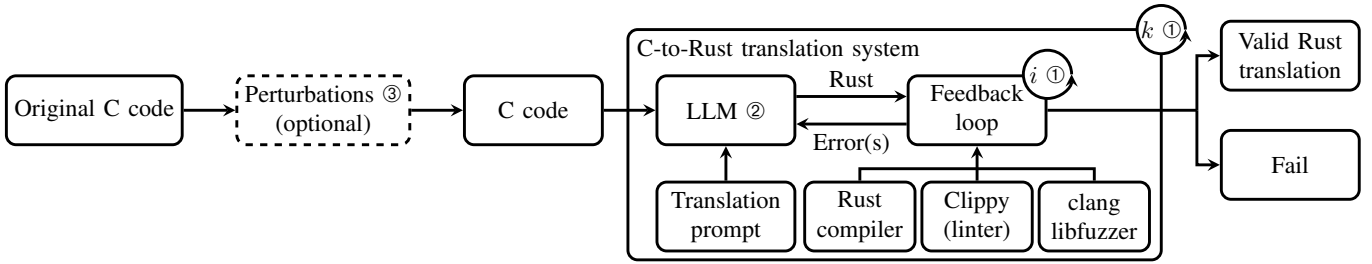

LLM-based systems can potentially make better use of the target language's natural concepts.
However, an LLM-based C-to-Rust translation system also needs to ensure that translated Rust code (\textit{i}) compiles without error and (\textit{ii}) is functionally equivalent to the original C~code.
For the former, the detailed error messages from the Rust compiler can be directly leveraged.
For the latter, our concrete translation system (see Fig.~\ref{fig:overview}) uses differential fuzzing~\cite{DBLP:journals/corr/abs-2405-11514}.
When these checks return a negative result, the LLM is automatically re-prompted. This feedback loop enables our translation system to repair generated output and to deliver results with quality guarantees. The latter is essential to enable using such techniques in a safety-relevant industrial environment.

We use that \textit{LLM-based \mbox{C-to-Rust} translation system} to investigate the influence of different variation points on translation success rates.
We aim to answer the following research questions:
\begin{itemize}
\item \textbf{RQ1:} How do internal feedback loops and multiple runs influence performance (\ie translation success rate)?
\item \textbf{RQ2:} How does LLM selection impact performance?
\item \textbf{RQ3:} How do variations of the input C~code (so-called \textit{code perturbations}) affect performance?
\end{itemize}
We perform the evaluation based on a dataset comprising both internal automotive embedded and external open-source C~code.
The results give insights into the performance, sensitivity, and robustness of an LLM-based translation system.
Our results show that LLM selection has an influence on performance (up to 9\% points difference), but feedback loops have an even bigger effect. 
Especially the first feedback loop has a high impact and can boost performance by up to 24\%. 
Using both feedback loops and multiple runs can boost performance by up to 50\%.
For most code perturbations, \gptnew\ turns out to be robust against them.

We provide the following main contributions:
\begin{enumerate}
	\item an evaluation of the influence of feedback loops and multiple runs on the success rate of an LLM-based \mbox{C-to-Rust} translation system,
	\item an evaluation of its performance across various LLMs, analyzing differences across models, and
	\item an analysis of the \textit{robustness} of the translation system, \ie performance under various code perturbations.
\end{enumerate}

The paper is structured as follows:
Sec.~\ref{sec:background} introduces background and related work,
Sec.~\ref{sec:translation-system} shows our translation system,
Sec.~\ref{sec:setup} contains the experimental setup,
Sec.~\ref{sec:results} presents the results,
Sec.~\ref{sec:discussion} discusses our findings, and
Sec.~\ref{sec:conclusion} concludes.

\section{Background and Related Work}
\label{sec:background}

We give an overview of generative AI for software engineering with a focus on the \textit{generate-and-check} pattern and its application in C-to-Rust translation.

\subsection{Generative AI for Software Engineering}

Generative AI and in particular text-based \textit{Large Language Models (LLMs)} have seen immense adoption in  software-engineering tasks~\cite{DBLP:journals/tosem/HouZLYWLLLGW24,fan_large_2023,rondon2025evaluatingagentbasedprogramrepair,alshahwan2024automated,DBLP:journals/corr/abs-2405-11514,DBLP:journals/corr/abs-2404-18852,zhang2024scalable}.
The progress in LLMs has been staggering with rapid progress on newly-created coding benchmarks such as SWE-Bench~\cite{jimenez_swe-bench_2024} and SWE-Lancer~\cite{miserendino2025swelancerfrontierllmsearn}.

In particular, LLMs are adopted in various \textit{LLM-based systems} for software engineering. Most prominently, we see this in AI-assisted code editors such as \copilot. 
In many tools, LLMs are a background AI engine that can be easily exchanged. 
They are used as text-based models that allow users to provide a textual task description and code as a \textit{prompt} and return a text \textit{response}. 
For an overview on LLMs especially from the viewpoint of software engineering, we refer to existing surveys~\cite{DBLP:journals/tosem/HouZLYWLLLGW24, fan_large_2023}. 
For the context of this work, we only need to consider that the used LLMs are types of machine learning models, typically based on deep learning~\cite{Goodfellow-et-al-2016}, transformers~\cite{vaswani2017attention}, and large-scale \mbox{(pre-)training}~\cite{abdin2024phi4technicalreport, huyen2025ai} such that they are capable to generate text across a wide range of tasks.

The representation of text for LLMs is a vector representation, called \textit{tokens}, and as such LLMs see text input and text output as a sequence of tokens.
More importantly, LLMs used in this work are sequence models, \ie they generate a sequence of tokens.
This is crucial, because in the sequential generation of tokens, there are various strategies to sample the next token from the distribution of potential next tokens~\cite{huyen2025ai}.
For the context of this work, the particular mechanism is not important, yet it is vital to understand that under different sequential samplings the same LLM may provide different responses for the same prompt.
A so-called \textit{temperature} is used to control variability, where higher temperature means more and a temperature of zero means no variability.

\subsection{Generate-and-Check Pattern}

Generative models and in particular LLMs have various desirable properties. 
First, they allow for flexible task formulation, and thus we can (in principle) use the same model across various tasks by only changing the task description in the prompt. 
Second, as discussed above, such models can \textit{generate} several responses, \ie possible solutions, to the same prompt.
A challenge in many tasks is that some responses may be incorrect and developers may (unintentionally) run into accepting wrong solutions.
However, this issue would be remediated if we could choose the ``best'' or a ``working'' solution. 
While such an approach is typically not helpful if the human has to perform the tedious and error-prone task of solution selection, it is very beneficial if there is an automated oracle that selects the ``best'' result.
Such oracles are particularly relevant when the output is a formal or semi-formal artifact such as code, and the task allows us to formulate and \textit{check} a property such as compilability of the code in the target language and behavioral invariance between C and Rust code.

In general, using such a \textit{generate-and-check} pattern requires an automated oracle to \textit{check} responses for their quality after \textit{generation}. 
Quality aspects typically include correctness, consistency, and completeness.
When we have such checks, we can leverage the power of feedback: 
If the quality does not suffice, feedback about failed checks can be fed back to the LLM to address and fix identified shortcomings. 
In these cases, we can create an automated feedback loop that helps to achieve certain qualities.

A concrete example for such a feedback loop in a code generation scenario is a compile check: The generated code must comply with the language, which can be checked by running a compiler. 
If compilation fails, the compiler provides error information and ideally hints how to fix it. 
Providing this information to the LLM helps it to fix the code and make it compilable~\cite{deligiannis2025rustassistant,DBLP:journals/corr/abs-2404-18852}. 
Our C-to-Rust translation system leverages feedback from the compiler and a behavioral equivalence check. 
We will see below that this so-called \textit{feedback loop} boosts the number of translation successes considerably.

As standard in code-related software tasks, our experimental evaluation leverages the \textit{pass@$k$} metric~\cite{DBLP:journals/corr/abs-2107-03374} for evaluation.
Since for several runs on the same problem, an LLM\,---\,and similarly our translation system\,---\,may return different results, a common approach is to evaluate the success of such models and systems across $k$ runs. 
Particularly, we perform a number of $n$ runs and evaluate whether at least one of $k$ runs produces an acceptable result, \eg compilable Rust code.
Running a higher number of experiments $n \geq k$ enables a more realistic statistical evaluation. For small $n$ (close to~$k$), we would otherwise underapproximate pass@$k$, especially for small $k$~\cite{DBLP:journals/corr/abs-2107-03374}.
We report $k$ and $n$ in our evaluations for each experiment. As standard in many works, we use $k=5$. For all basic experiments without perturbations, we leverage $n=20$.

\subsection{C-to-Rust Translation}

Creating idiomatic Rust code that behaves equivalently to existing C code is a challenging problem because there exist language features such as global variables that are only present in one of the two languages.
The classical approach to translate code from C to Rust is to use a syntax- and rule-based system~\cite{c2rust, DBLP:journals/pacmpl/EmreSDH21, DBLP:journals/corr/abs-2412-15042, DBLP:journals/pacmpl/HongR24}.
Such approaches typically do not exploit Rust's strengths and often result in unreadable code that will be difficult to maintain.
In contrast, LLM-based approaches have the potential to consider the idioms of the target language and thus produce better code.

When using LLMs for C-to-Rust translation, the first goal is to obtain compilable Rust code.
The LLM can iterate on error message by the Rust compiler to obtain compilable Rust code~\cite{DBLP:journals/corr/abs-2405-11514, DBLP:journals/corr/abs-2404-18852, DBLP:journals/corr/abs-2409-10506, DBLP:journals/ese/HongR25} and we leverage this approach in our translation system.
Previous work describes two directions to obtain compilable Rust code with equivalence indication between C and Rust code.
The first approach is to compile C (or other major languages) to Web Assembly and then use the tool \texttt{rWasm} to obtain (unreadable) Rust code and use this as an oracle program.
An LLM can then suggest (readable) Rust code that is first checked for compilability and then for equivalence with the oracle program with the possibility for feedback loops~\cite{DBLP:journals/corr/abs-2404-18852}.
The second approach also uses an LLM to suggest Rust code and then makes it compilable.
Afterwards, differential fuzzing between the C and the Rust code tests input/output equivalence for random inputs within a given time limit~\cite{DBLP:journals/corr/abs-2405-11514}. 
JSON serialization is used to map C and Rust values for differential fuzzing in that work.
We also leverage differential fuzzing with automated harness generation for (weak) equivalence checking. In contrast to previous work, we use explicit type conversion, as it gives more control over which values should (not) be mapped to each other.

\subsection{Robustness}
\label{sec:background_robustnesss}

While robustness evaluation has also been studied in other contexts such as natural language processing~\cite{DBLP:conf/aaai/MalfaK22, DBLP:journals/access/OmarCNM22, DBLP:conf/naacl/AhujaAGWSOHJABS24} or image recognition \cite{DBLP:journals/corr/abs-2209-02408, DBLP:conf/iccvw/MullerBK23}, we focus on robustness evaluation in code generation.
Wang \etal present a benchmark designed to assess the robustness of code generation models~\cite{wang_recode_2023}.
They introduce metrics designed to capture wort-case performance across perturbed inputs but focus on docstring-based code generation and code completion for Python.
One of their metrics used in this work is \textit{robust pass@$k$}.
It extends pass@$k$ by incorporating worst-case performance over $s$ randomly perturbed prompts.
Mastropaolo \etal examined the robustness of \copilot\ by analyzing how semantically equivalent natural language instructions affect code generation~\cite{mastropaolo_robustness_2023}.
They use deep learning, heuristics, and manual work to obtain perturbed descriptions and focus on function generation based on method signatures and Javadocs.
Yan \etal concretize instructions based on features of the code, \eg loops or function definitions~\cite{YAN2025107645}.
The concretizations also define if code features should be integrated or left out.
Improta \etal analyze the robustness of offensive code generators~\cite{improta_enhancing_2025}, \eg used for penetration testing.
They introduce the two perturbation strategies word substitution and word omission in a way that preserves the original meaning of the instruction.
There also exists work on adversarial robustness where perturbations are designed in an attempt to mislead the model~\cite{michel_evaluation_2019, li_textbugger_2019, cheng_seq2sick_2020, wu_evaluating_2020, boucher_bad_2022}.
Our work leverages several of these perturbations plus some newly created ones (\cf Sec.~\ref{sec:perturbations}).

\section{C-to-Rust Translation System}
\label{sec:translation-system}

Fig.~\ref{fig:overview} depicts our translation system, which is inspired by previous work~\cite{DBLP:journals/corr/abs-2405-11514, DBLP:journals/corr/abs-2404-18852, DBLP:journals/corr/abs-2409-10506, DBLP:journals/ese/HongR25}. 
It uses an LLM as translation engine and different tools to check the LLM's output. 
If the LLM makes a translation mistake, the LLM is re-prompted to fix its mistake.
The input to the system is C~code that is passed to the LLM with the following translation prompt:  ``\textit{Translate the following C~code to Rust. Keep all identifiers exactly as they are. $<$C\_code$>$}''.
Note that the second sentence of the prompt increases the chances of successful differential fuzzing later on.
The LLM produces an answer for the prompt from which we extract the Rust code.
We decided to use a simple prompt as a baseline, as our focus is on other variation points. Future work could experiment with prompts as another factor impacting performance.

Next, we check the Rust code for correctness and equivalence to the original C~code.
First, we try to compile the Rust code with the Rust compiler~\texttt{rustc} and use~\texttt{clippy} for linting.
Errors reported by both tools are fed back to the LLM with the following prompt ``\textit{You made the following mistakes: $<$error\_messages$>$}''. As the Rust compiler gives detailed error messages, this is often helpful for the LLM to fix its mistakes.

When the Rust code compiles and is linted successfully, the equivalence between the C~code and the Rust code is checked.
Here, we use differential fuzzing with a fixed timeout.
We use the builtin fuzzer of \texttt{clang} (libFuzzer\footnote{\url{https://llvm.org/docs/LibFuzzer.html}}) and call the functions of the C~code and Rust code from a generated C~code harness.
For the latter, we analyze the interface of the original C~functions \wrt parameters, return types, and global variables and generate a suitable harness that initializes inputs (from the raw data provided by the fuzzer), calls both functions, and compares outputs. The Rust function is called via Rust's foreign function interface (FFI) and includes a type mapping and conversion between C and Rust types, which is also generated automatically.
The harness and original C~function are compiled to an executable along with clang's fuzzer, and the Rust function is linked in as a static library. Then, the fuzzer is executed. As the fuzzer terminates not only on different output, but also when a runtime error occurs, we only consider it to be relevant if the error occurs only in the Rust code, but not in the C~code. Otherwise, fuzzing is continued.

Differential fuzzing either finds a counterexample, \ie different behavior between the C and the Rust code, or reaches the timeout without finding such a counterexample.
Found counterexamples are fed back to the LLM with the same prompt as above.
In this work, not finding a counterexample within the time budget is used as indication for equivalence between the C and the Rust code and deemed a translation success. Although it is not a proof of equivalence, it is still a strong indication for it, given the thorough exploration of the input space by the fuzzer and the rather small size of the benchmark functions that we use in our experiments.
Stronger checks towards a formal equivalence proof could be used in the future for a productive system but are out of scope for this experimental study.

\section{Experimental Setup}
\label{sec:setup}

This section details on the experimental setup: the benchmark, chosen LLMs, code perturbations, and the procedure.

\subsection{Characterization of the Benchmark}
\label{sec:char-benchmark}

Our benchmark consists of 50~C~files with 76~C~functions.
Thirty of the files are internal automotive embedded code~\cite{DBLP:conf/vmcai/HeckingHarbuschQS24}, ten files are from open-source code (also used in previous work~\cite{DBLP:journals/corr/abs-2405-11514}), and ten files come from competitive programming solutions (also used in previous work~\cite{DBLP:journals/corr/abs-2404-18852, DBLP:conf/iclr/SzafraniecRLLCS23}).
Due to the inclusion of internal code, we are unable to release the full dataset.
Keeping code internal has the benefit that LLMs should not have seen it during training.
All three sources of our benchmark originally consist of more than the used files.
We sample from the set of all files to get a manageable number of files for our experiments.
We use files across benchmarks to identify differences between automotive embedded code and typical open-source benchmarks to better gauge how reported results in literature would translate to our domain.

We give more details for each of the three groups included in our benchmark.
The automotive code includes functions from different domains and different layers in the AUTOSAR Classic architecture~\cite{autosar}. It contains low-level code, base software library code, and application-level control code. The largest share is hand-written code, but some generated files are also included (\eg from AUTOSAR RTE).
The open-source code originates from two C libraries, one for audio processing and one for sound card emulation. 
The competitive programming examples originate from a large scale data scrap used for unsupervised learning.
For more information about the open-source code, we refer to the original papers.
Table~\ref{tab:metrics} shows metrics of the individual C files from our benchmark.
Overall, the files are small to medium in size (up to 296~lines of code). The table shows the overall metrics and the numbers for the internal files. Notably, the internal files are slightly smaller on average but have more control flow branches (CC).

\begin{table}
\caption{Characteristics of benchmark files. NLOC excludes comments and empty lines; tokens is the number of tokens for \gptold; CC is the average cyclomatic complexity of functions.}
\label{tab:metrics}
\centering
\begin{tabular}{l|rrrr|rrr}
\toprule
& \multicolumn{4}{c|}{All Files} & \multicolumn{3}{c}{Internal Files} \\ 
\midrule
Metric & min &   avg & std.\,dev &  max &    avg & std.\,dev &  max \\ 
\midrule
LOC    &  15 &  40.0 &    45.2 &  296 &   36.8 &    29.1 &  158 \\ 
NLOC   &   7 &  28.9 &    38.1 &  249 &   25.9 &    22.7 &  135 \\ 
Tokens &  48 & 454.2 &   814.2 & 5716 &  417.9 &   345.6 & 1696 \\ 
CC     &   1 &   2.4 &     2.3 &   13 &    4.3 &     4.2 &   13 \\ 
\bottomrule
\end{tabular}
\end{table}

\subsection{Used Models}

We compare three very different models to study their impact on the C-to-Rust translation system: 
(\textit{i})~An old, powerful model\,---\,namely \gptold~\cite{openai2023gpt35turbo}\,---\,that was used extensively in previous work. It serves as a reference to compare against and see progress of recent (small) models. 
(\textit{ii})~A standard, commercial model\,---\,namely \gptnew~\cite{openai2024gpt4omini}\,---\,that is in widespread use due to its interesting cost-performance and availability characteristics. 
(\textit{iii})~A recent open model\,---\,namely \msphi~\cite{abdin2024phi4technicalreport}\,---\,as open models seem to track the performance of closed commercial models with some time delay, allow an even better cost-performance trade-off, and may be used locally by developers. All models in this paper are used with their default parameters, and we set the temperature to 0.7. We performed experiments outside the scope of this paper that showed that the temperature's influence is negligible for a wide range of values in common models.

\subsection{Perturbations}
\label{sec:perturbations}

Our \textit{prompts} for the LLM consist of a simple \textit{instruction} describing the translation task and C \textit{code} to be translated~(\cf Sec.~\ref{sec:translation-system}).
The C code can contain \textit{comments}.
We apply perturbations to code and comments and want to evaluate robustness in real-world scenarios and not in adversarial settings.
Therefore, all perturbations need to ensure syntactical correctness, semantic equivalence, and real-world relevance.
We also differentiate between deterministic and stochastic perturbations because some perturbations deploy randomized strategies.
We strive for variety of perturbations and thus leverage previous work on levels of program transformations~\cite{faidhi_empirical_1987}.
The six levels of program transformation that we tackle are (I) \textit{comments \& indentation}, (II) \textit{identifiers}, (III) \textit{declarations}, (IV) \textit{functions}, (V) \textit{semantic equivalents}, and (VI) \textit{decision logic \& expressions}.
The goal is to sample perturbations across levels to avoid biasing the results by overemphasizing particular kinds of perturbation.

On Level I, we use the existing perturbation strategies of round-trip translation (to and from German) and typo insertion for comments~\cite{wang_recode_2023} as well as comment removal~\cite{zhang_challenging_2023}.
We add new perturbation strategies inserting further comments using an LLM and using existing formatters on the code indentation.
On Level II, we use typo insertion and changes of naming conventions for identifiers~\cite{wang_recode_2023} as well as replacing identifiers with short ones like $a$, $b$, $c$, and so on~\cite{yu_data_2022}.
We newly use round-trip translation (to and from German) also for identifiers and add further changes to identifier naming conventions as well as trying to improve them with an LLM.
On Level III, we use constant insertion~\cite{cheers_spplagiarise_2019} and dead code insertion~\cite{wang_recode_2023, zhang_challenging_2023} and add the insertion of function declarations for existing includes in the C code.
On Level IV, we use the extraction of code into functions using an LLM~\cite{cheers_spplagiarise_2019} and add changes to function signatures.
On Level V, we use swapping of {\tt for} and {\tt while} loops~\cite{wang_recode_2023, cheers_spplagiarise_2019, li_exploring_2024, zhang_challenging_2023, yu_data_2022} and of conditions~\cite{wang_recode_2023, li_exploring_2024, yu_data_2022}.
On Level VI, we use condition duplication~\cite{li_exploring_2024} and add an application of DeMorgan's law.

\begin{table}[t]
	\caption{Counts of perturbations by level}
	\label{tab:perturbation_counts}
	\centering
	\begin{tabular}{l|cr|c}
		\toprule
		Level & \multicolumn{2}{l|}{Previous Work} & New \\
		\midrule
		I & 8 & \cite{mastropaolo_robustness_2023}, \cite{wang_recode_2023}, \cite{zhang_challenging_2023} & 6 \\
		II & 8 & \cite{wang_recode_2023}, \cite{yu_data_2022} & 8 \\
		III & 6 & \cite{cheers_spplagiarise_2019}, \cite{wang_recode_2023} & 3 \\
		IV & 3 & \cite{cheers_spplagiarise_2019} & 3 \\
		V & 2 & \cite{cheers_spplagiarise_2019}, \cite{li_exploring_2024}, \cite{wang_recode_2023}, \cite{yu_data_2022}, \cite{zhang_challenging_2023} & 0 \\
		VI & 3 & \cite{li_exploring_2024} & 1 \\
		\midrule
		Total & 30 & & 21 \\
		\bottomrule
	\end{tabular}
\end{table}

Table~\ref{tab:perturbation_counts} summarizes the used perturbations.
We use 30 perturbations from the aforementioned related works and 21 newly created ones. 
Our newly created perturbations lead to a better distribution of perturbations across levels~\cite{faidhi_empirical_1987,cheers_spplagiarise_2019}.

\subsection{Procedure}

We run our C-to-Rust translation system on all 50~files of the benchmark. If there is no acceptable result after at most five iterations of the inner feedback loop, the translation has failed; otherwise, it was successful. During the translation process, we track in which iteration which quality stage (\ie compilable, linted, or fuzzed) was reached. Where not otherwise stated, a translation is only considered successful if differential fuzzing does not find any difference in behavior (``final result''). This also implies that all previous checks were successful. The whole process is run 20~times to allow for more representative pass@$k$ calculations. 
For \textbf{RQ2}, we repeat these runs using different LLMs. 
For \textbf{RQ3}, we additionally perform different perturbations on the input C~code.

\section{Results}
\label{sec:results}

In the following, we discuss our experimental results \wrt our three research questions.

\subsection{\textbf{RQ1}: How do internal feedback loops and multiple runs influence performance?}

\begin{table}
\caption{Pass@5 Results Across Iteration Counts}
\label{tab:pass_at_k}
\begin{tabular}{lccccc}
\toprule
Result Type & $\leq\ $ 1 & $\leq\ $ 2 & $\leq\ $ 3 & $\leq\ $ 4 & $\leq\ $ 5 \\
\midrule
Compilation success & 0.88 & 0.98 & 0.98 & 1.00 & 1.00 \\
Final result & 0.69 & 0.78 & 0.78 & 0.78 & 0.79 \\
\bottomrule
\end{tabular}
\end{table}


In Table~\ref{tab:pass_at_k}, we show pass@$5$ for 20~experiments with \gptnew\ for numbers of LLM invocations~$i$ from at most one to at most five. 
Results are shown both for compilation and fuzzing success. First, we can see that\,---\,as expected\,---\,compilation is a simpler task to achieve and reaches almost perfect success when leveraging feedback loops. 
Second, there is a substantial drop to a successful translation (\ie including fuzzing). 
Third, we can see that without a feedback loop, \ie iteration count of at most one, there is a substantially reduced performance when compared to the use of a feedback loop (\ie iteration count of at most two or more). 
A single feedback loop iteration increases compilation and fuzzing success significantly. With further iterations of the loop (which become more involved and expensive), there are only slight performance improvements, such that from a price performance perspective, a single additional feedback loop seems like a practical approach.

\begin{figure}[t]
	\centering
	\includegraphics[width=\linewidth]{code/results/GPT-4o-mini/pass_at_k_plot_tokens.pdf}
	\caption{pass@$k$ performance versus sum of generated tokens: The curves show how solution success rates (from pass@$1$ to pass@$5$) improve with larger token budgets for increasing maximal iterations of inner feedback loop.}
	\label{fig:gpt4o-tokens}
\end{figure}

Fig.~\ref{fig:gpt4o-tokens} shows how cost in terms of token counts relates to different iteration counts $i$, repetitions $k$, and thus performance. 
We can see that in all cases, it pays off to run the inner loop at least once. 
This diminishes with $i$ at most two or more, and there it might often make more sense to just perform a new run of the complete translation system and thus increase $k$. 

\begin{figure}[t]
	\centering
	\includegraphics[width=\linewidth]{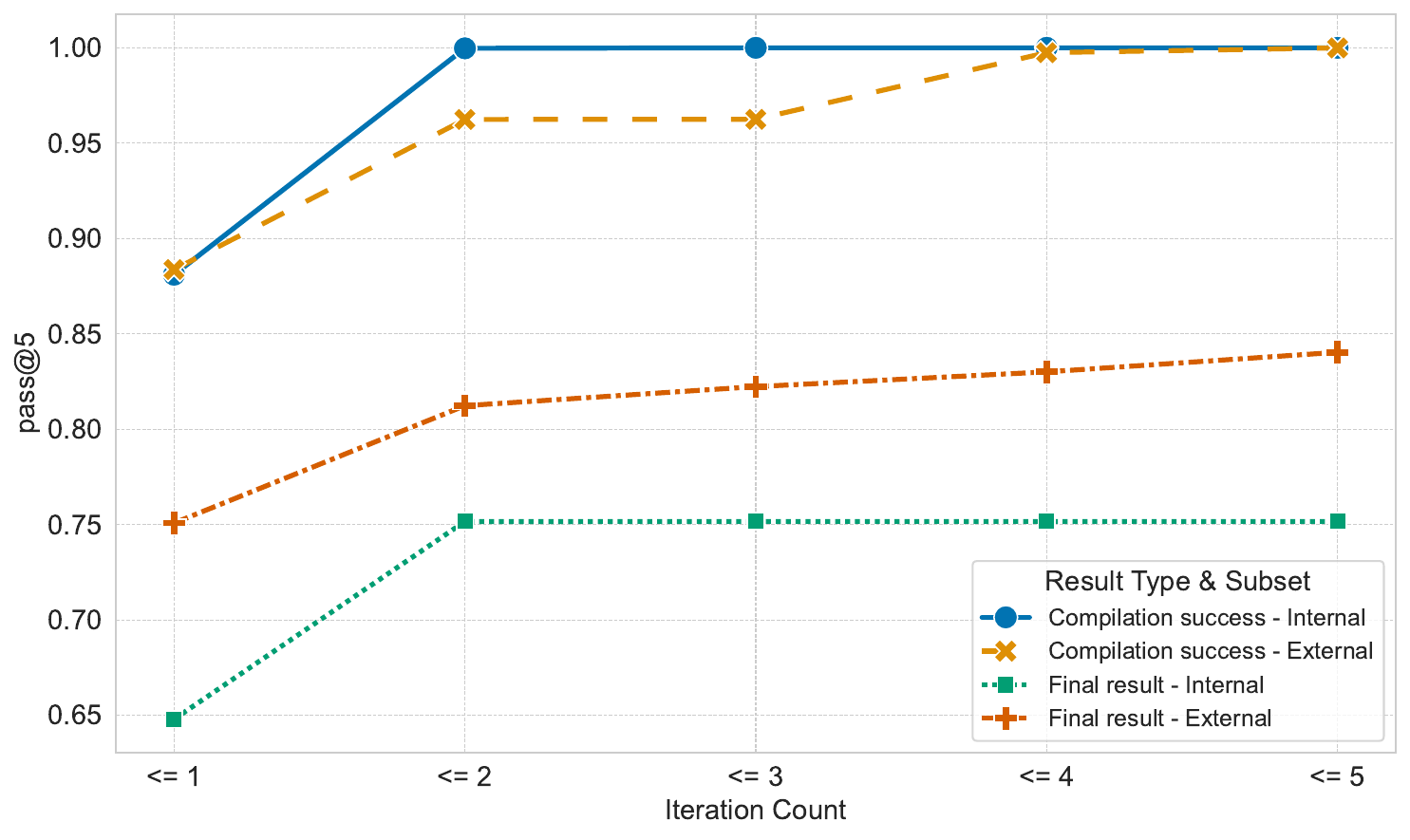}
	\caption{pass@$5$ performance: Comparing internal benchmark files versus external benchmark files.}
	\label{fig:gpt4o-internal-external}
\end{figure}

To check for differences between internal and public files in our benchmark, we compare results across internal and external files in Fig.~\ref{fig:gpt4o-internal-external}. 
It shows that our translation system has a 12\% higher performance for external files in comparison with internal files for final result and at most five internal feedback loops. 
We also observe that the inner loop only has an effect for the first iteration on internal files, while it still has a positive effect on external files. This might be due to the differences in length (\cf Sec.~\ref{sec:char-benchmark}).

\begin{figure}[t]
	\centering	\includegraphics[width=\linewidth]{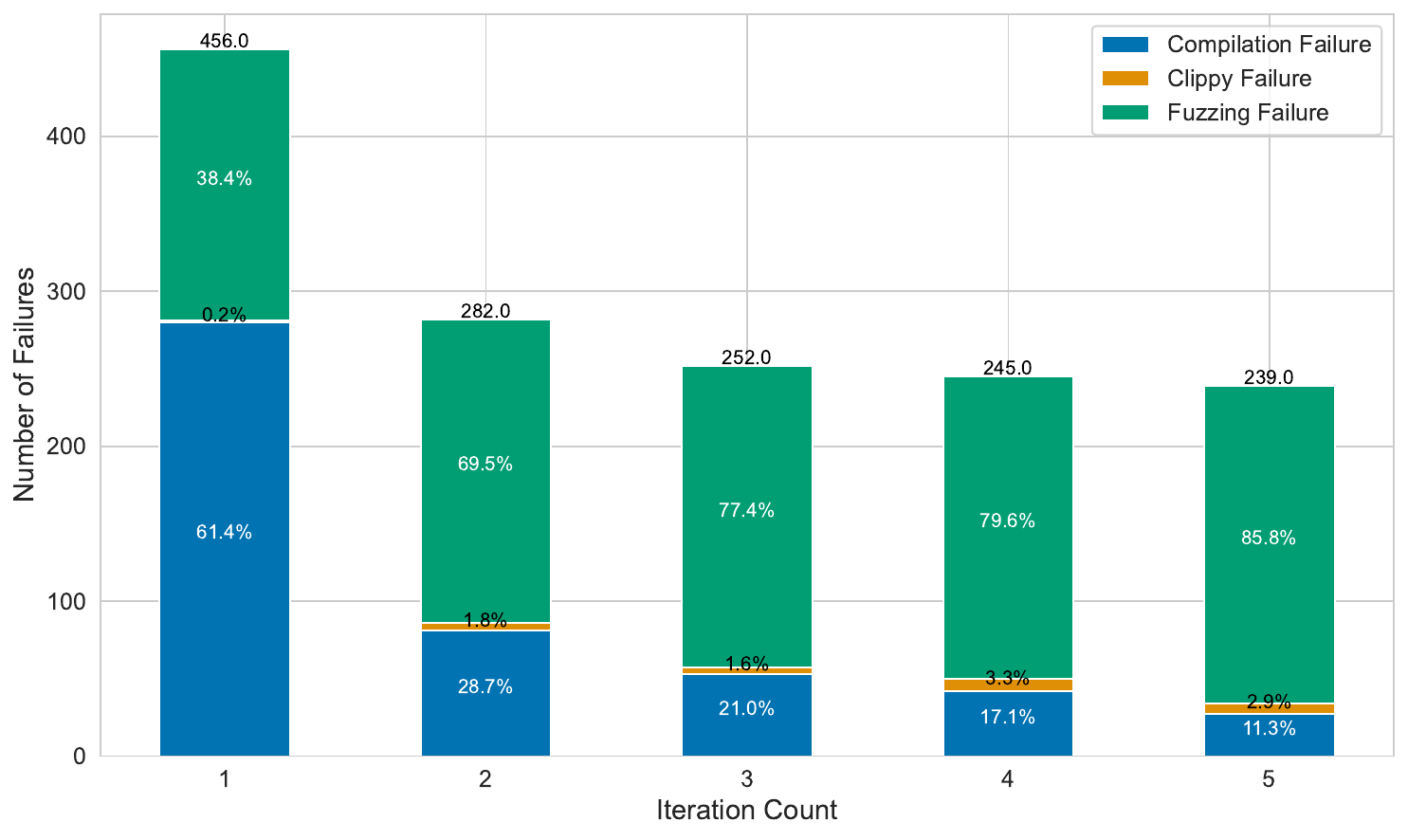}
	\caption{Detailed analysis of which check fails the translation tasks for each iteration of the feedback loop.}
	\label{fig:gpt4o-innerloop-errors}
\end{figure}

Let us detail on why translations fail. Fig.~\ref{fig:gpt4o-innerloop-errors} summarizes fails across all 20~runs with \gptnew. 
We can see the absolute counts which decrease over iterations, since we only run an additional iteration for previous fails. 
Again, we can see a big difference between single iterations and ones with feedback: 
For a single iteration, we may run into substantial compilation problems, but these mostly get resolved in the first feedback loop. There is continuous further reduction but much less notable than after the first iteration.

\begin{framed} \noindent
\textbf{Findings:} The feedback loop significantly improves the translation success rate. The first loop iteration brings most benefit, up to 24\%, while subsequent ones lead to minor improvements. In combination with repetitions, the feedback loop helps to achieve highest performance at additional cost.
\end{framed}

\subsection{\textbf{RQ2}: How does LLM selection impact performance?}

\begin{figure}[b]
	\centering
	\includegraphics[width=\linewidth]{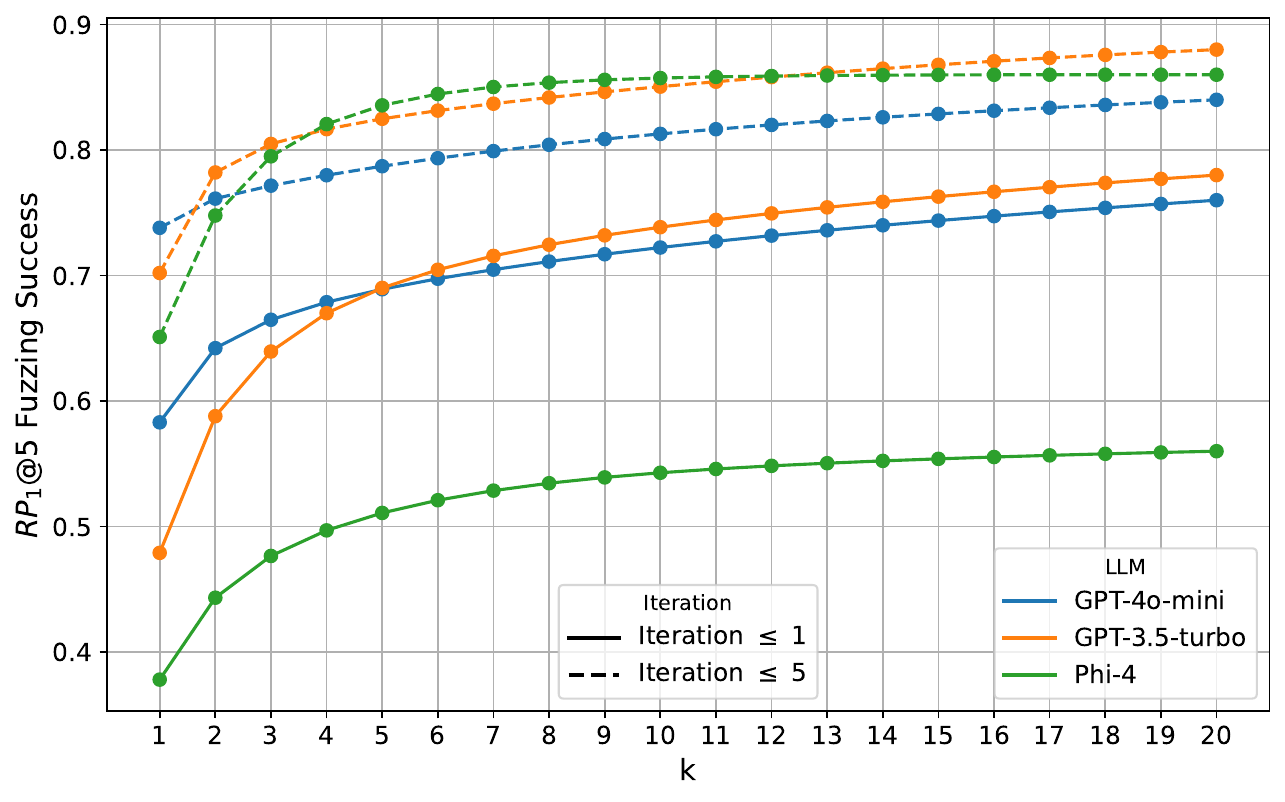}
	\caption{pass@$k$ performance across three models for $k$ up to 20 runs. Different styles are used for varying iteration counts.}
	\label{fig:model_comparison}
\end{figure}

Fig.~\ref{fig:model_comparison} shows the results of running our translation benchmark on three models: \gptold, \gptnew, and \msphi.
Here, we measure the pass@$k$ metric for different~$k$.
We can see three interesting points in the plot: 
(\textit{i})~For single iterations (solid lines), the two GPT models clearly outperform \msphi. 
(\textit{ii})~For at most five iterations (dotted lines), \msphi~performs better than the two GPT models in the region of four to eleven runs~$k$.  
(\textit{iii})~While \msphi~plateaus, the two GPT models continuously improve with~$k$, showing that these models seem to have additional capacity for ``creativity'' in the sense of generating new solutions.
Additionally, let us have a look at the difference between the dotted lines (at most one iteration) and the solid lines (at most five iterations). 
Across all models, we see a solid boost using a feedback loop. 
This is especially  true for \msphi, which goes from clearly worst model to best model in the aforementioned region of four to eleven runs~$k$.

\begin{figure}[t]
	\centering
	\includegraphics[width=0.6\linewidth]{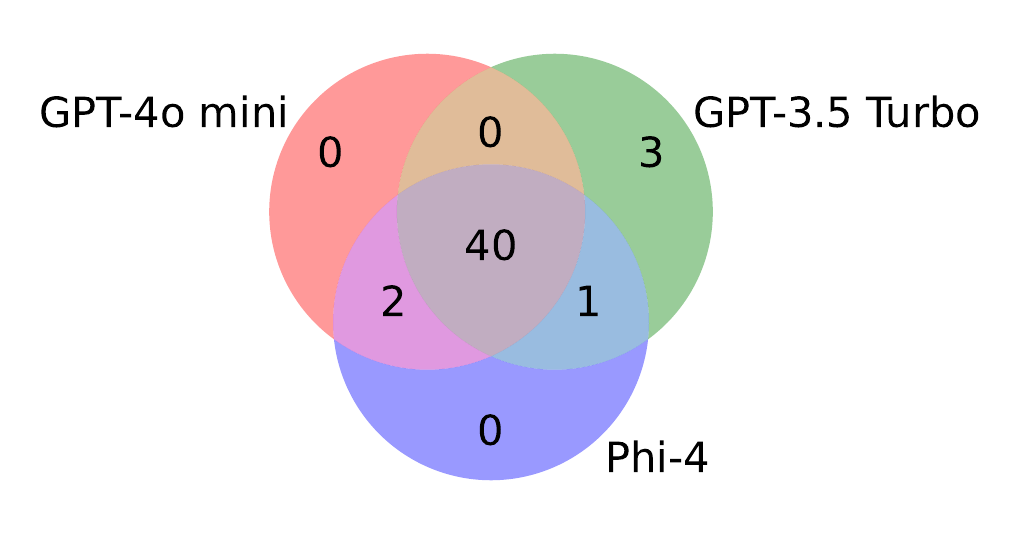}
	\caption{Number of files that are successfully translated by different models (by at least one of 20~runs with 5~iterations each).}
	\label{fig:solved_comparison}
\end{figure}

Let us now compare which files are correctly translated by each model across all runs performed. 
The Venn diagram in Fig.~\ref{fig:solved_comparison} shows that the different models mostly solved the same programs, but there are some differences that could be leveraged for increasing the solve rate even further. 
Overall, we can see that across models, we can translate 46~programs, while the best model can translate only 44~files (\gptold). 
Note that the absolute numbers per model correspond to pass@$20$ in Fig.~\ref{fig:model_comparison} for the 50~files in our benchmark and the number across all three models corresponds to pass@$60$ of $0.92$.

\begin{framed} \noindent
\textbf{Findings:} There are significant differences between LLMs without a feedback loop: \gptnew\ is $\approx$ 50\% better than \msphi\ for pass@$1$. Feedback loops remediate the differences to a mere 13\%.
Moreover, when using a combination of \mbox{$k\in[4,11]$} with feedback loops, \msphi\ becomes the best evaluated model.
Lastly, we can increase performance at additional cost until models plateau for high $k$ ($\approx$ 10 for \msphi, but $> 20$ for both evaluated GPT models).
\end{framed}

\begin{figure}[t]
	\centering
	\includegraphics[width=\linewidth]{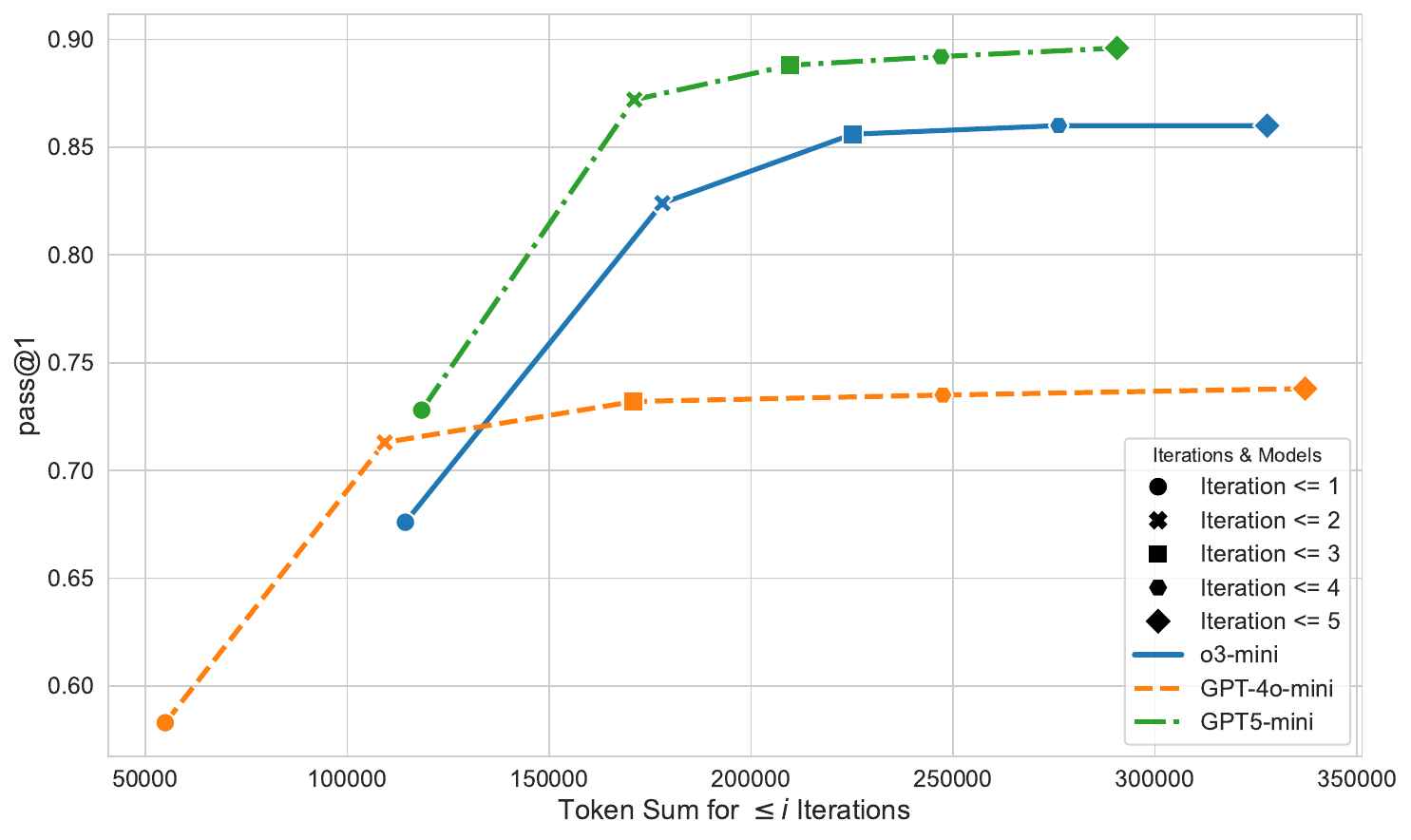}
	\caption{pass@$1$ performance and iterations over token count for different reasoning and non-reasoning models.}
	\label{fig:reasoning}
\end{figure}

Our main study focus are feedback loops.
We focus on classical non-reasoning models, since they do not include (hidden) reasoning. Thus, we see a direct effect of feedback loops.
Nevertheless, we obviously are interested whether such results we see there also hold for reasoning models. 
To this extent, we revisit Fig.~\ref{fig:gpt4o-tokens}, \ie a comparison of token effort versus performance, but particularly focus on pass@$1$ across different models for the inner feedback loop.
Additional to the pass@$1$ curve of \gptnew, we show the corresponding curves for reasoning models o3-mini~\cite{openai2025o3mini} and GPT-5 mini~\cite{openai2025gpt5mini}\footnote{For the reasoning models, we ran five experiments each.}.
As before we sum up all tokens, which for these models also include reasoning tokens.
We see several interesting differences in Fig.~\ref{fig:reasoning}: 
First, we can see that reasoning models spend additional tokens on the first iteration (without feedback), hence they are shifted to the right.
We can see that this pays off, as the out-of-the-box performance of these models increases (hence the shift upwards). The newer the model, the better it fares.
Second, we can see that the trend of improvement with the inner feedback loop is consistent across models.
We can see a big benefit of the first feedback, which subsequently diminishes.
Third and most subtle is that the token differences across models reduces and for the last iteration, the most tokens are even spent for \gptnew. This is because the better models do not actually start additional feedback rounds on files that they already successfully translated and thus use fewer tokens in additional iterations.

\begin{framed} \noindent
	\textbf{Findings:} The inner feedback loop significantly improves the translation success rate across all types of evaluated models whether non-reasoning or reasoning. Particularly, the first feedback provides a substantial boost with $\approx$ 20\% improvement across all evaluated models.
\end{framed}

\subsection{\textbf{RQ3}: How do code perturbations affect performance?}

\begin{figure}[t]
	\centering
	\includegraphics[width=\linewidth]{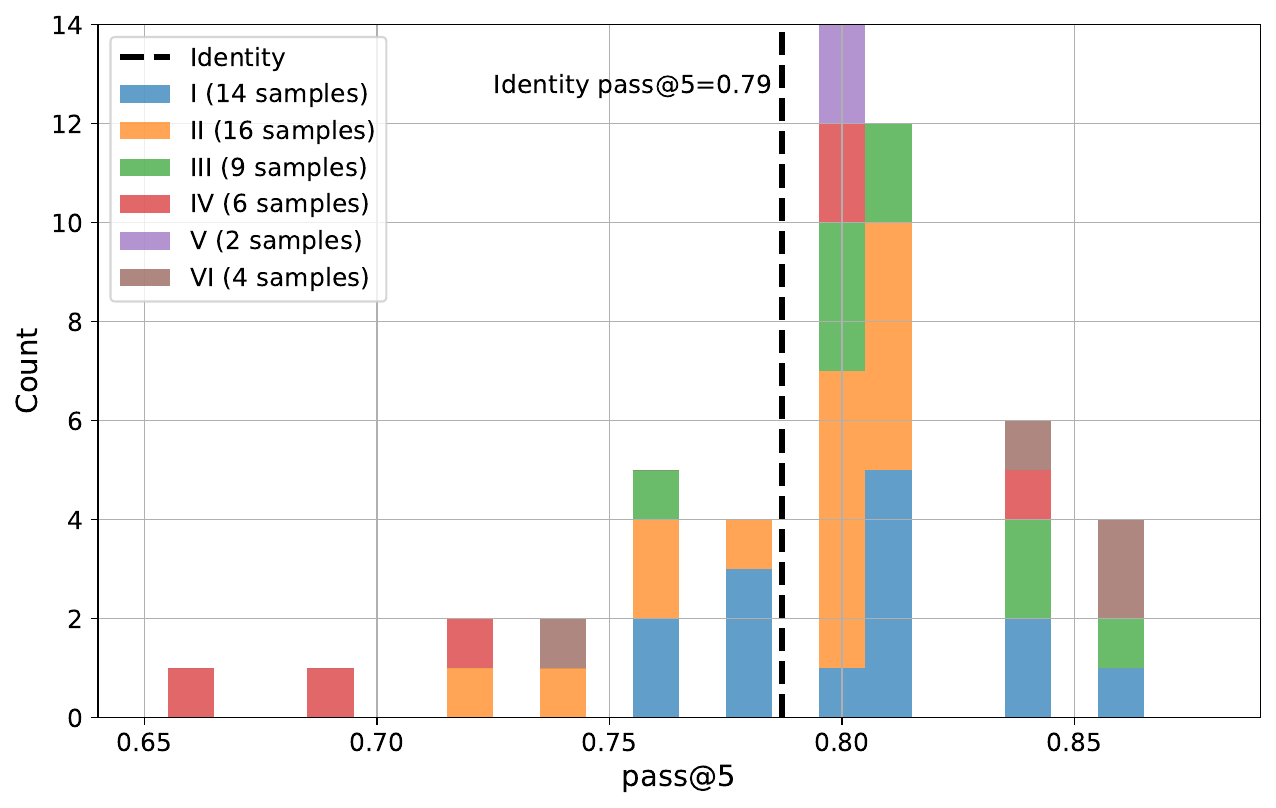}
	\caption{Histogram of pass@$5$ for all runs on \gptnew~based on level association. Mean Identity results as a dotted line for reference.}
	\label{fig:robuststochastic}
\end{figure}

Let us study code perturbations and their impact on translation system performance. 
Fig.~\ref{fig:robuststochastic} depicts a histogram of all runs for \gptnew\ across all 51 perturbations associated to levels I to VI with six different colors. 
The mean of runs without a perturbation, so-called \textit{Identity}, is indicated by the dotted line.
Perturbations from most levels concentrate around that line.
Level IV appears particularly challenging, which is due to a single perturbation: LLMCodeExtraction. This perturbation uses an LLM to extract code blocks into functions. 
We hypothesize that this perturbation adds more complexity to the control flow, leaving more room for errors in the translation.
For the sake of brevity, we omit our analysis where we compare LLMCodeExtraction against no perturbation (Identity) that supports this hypothesis.
Overall, most perturbations have little effect on the results compared to runs without perturbations, showing that \gptnew\ is robust against most of our selected perturbations.

As a follow-up experiment, we randomly sample sets of 20~perturbations (incl.~Identity). We construct 10,000 of these random perturbation set samples. 
Now, we aggregate the pass@$k$ values across the perturbations. First, we use the minimum value corresponding to a robust pass@$k$ as introduced in Sec.~\ref{sec:background_robustnesss}. 
This is a pessimistic view, looking at the worst case. 
In the original work~\cite{wang_recode_2023}, this is warranted for use cases where we cannot check the result for correctness.
However, for our translation system, we can check the results and select the ``best'' out of all samples. 
Thus, second, we also compute an optimistic view by using the maximum across all perturbations for each file.
This best case leverages diversity of results using \textit{data augmentation}~\cite{yu_data_2022, huyen2025ai}.

\begin{figure}[t]
	\centering
	\includegraphics[width=\linewidth]{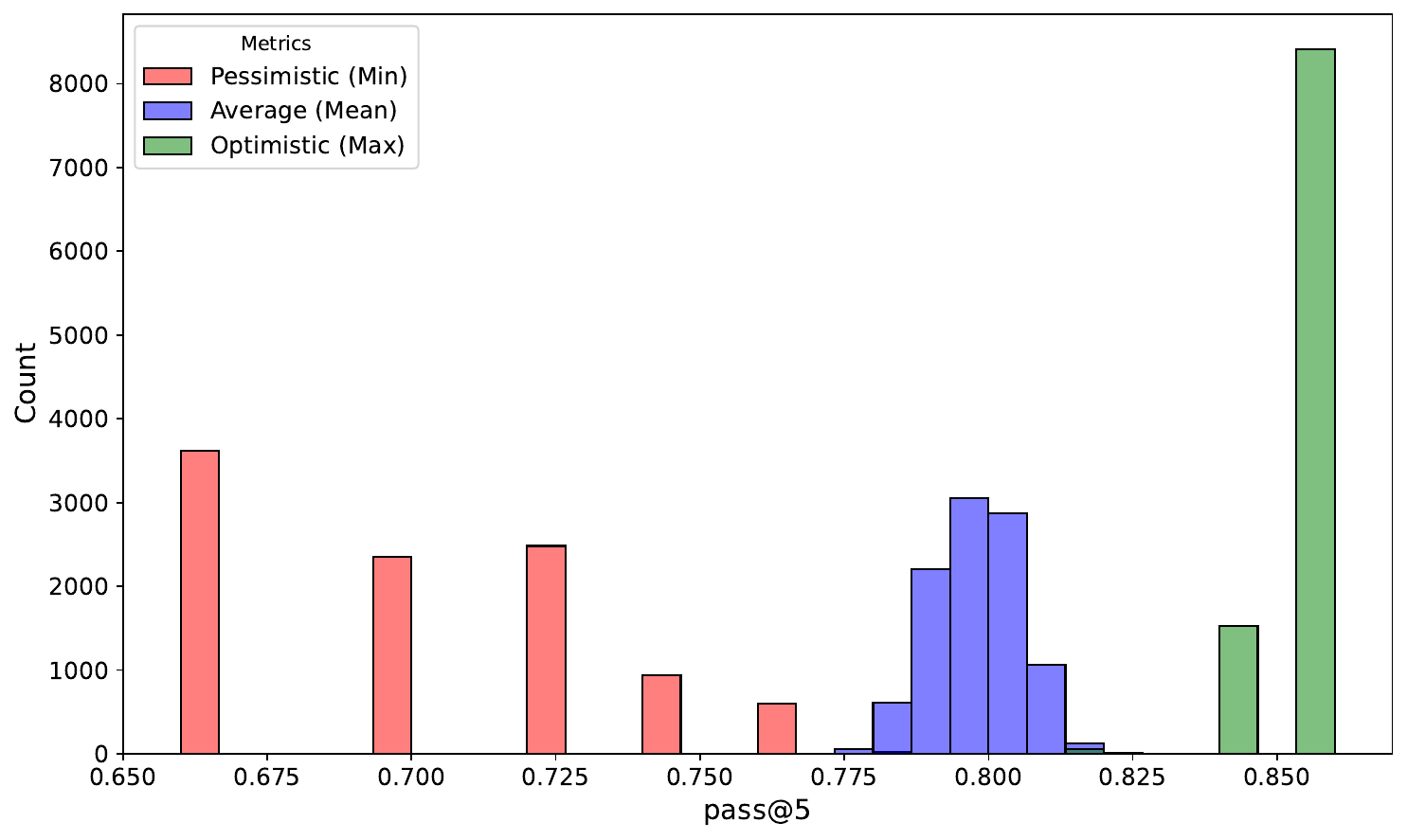}
	\caption{pass@$5$ for 10,000 random samples of 20 translation runs: Distribution of min, mean and max pass@$5$.}
	\label{fig:augmentation}
\end{figure}

Fig.~\ref{fig:augmentation} shows a histogram summarizing the pessimistic and optimistic results and provides an average view (computing mean pass@$k$ across all perturbations) as a reference. 
As we can see, there is a wide gap between the pessimistic and the optimistic view.
This gap underlines the value of an automated oracle, since only such an oracle allows us to transfer from a pessimistic performance level to an optimistic one.

Finally, note that perturbations particularly help when using the feedback loop.
When we sample 20 random perturbations, the mean pass@$5$ across 10,000 such samples is basically the same as for Identity, when we use five iterations.
However, without iterations, we see a small decrease in pass@$5$ (by~0.01). This shows again the benefit of using feedback.

\begin{framed} \noindent
\textbf{Findings:} \gptnew\ is robust against code perturbations across almost all hierarchy levels.
Diversity of perturbations can be leveraged for data augmentation, improving pass@$5$ performance up to 9\% from Identity runs.
\end{framed}

\section{Discussion}
\label{sec:discussion}


\subsection{Performance vs. C Code Characteristics}


In our experiments, we observed that the translation task almost always succeeded for small files with low token count, while larger files remain a challenge.
Files with up to 500~\gptnew\ tokens almost always succeeded, files with 500 to 1000~tokens sometimes succeeded, and files with more than 1000~tokens failed in most cases.
One large function is currently so complex that even reaching compilability can be seen as achievement. 
How to deal with large, complex functions in terms of functional equivalence remains as future work.

\subsection{Detailed Error Analysis}
\label{sec:detailederror}

In a few cases, the translation system fails. Table~\ref{tab:error_distribution} provides an overview of error cases across all experiments. 
Some errors occur due to issues in the fuzzing setup, \eg when some perturbations hinder the integration of the target program into the fuzzing harness. This is different for the second category, where we see an exception during the actual fuzzing runs. There are also a few errors in the translation system itself. Finally, we see issues when interfacing with LLMs (and LLM services). 
Note that such issues in translation, especially \wrt fuzzing setup, are not uncommon, and higher numbers are reported in previous work~\cite{DBLP:journals/corr/abs-2405-11514}.

\begin{table}[t]
	\caption{Distribution of error cases across different experiments. Largest errors per column are marked in \textbf{bold}.}
	\label{tab:error_distribution}
	\centering
	\begin{tabular}{l|c|c|c|c}
	\toprule
	\begin{tabular}{@{}c@{}}Experiment\\~\end{tabular} & \begin{tabular}{@{}c@{}}Fuzzing\\Setup\end{tabular} & \begin{tabular}{@{}c@{}}Fuzzing\\Exception\end{tabular} & \begin{tabular}{@{}c@{}}Translation\\System\end{tabular} & \begin{tabular}{@{}c@{}}LLM\\API\end{tabular} \\ 
	\midrule
	Identity & 0.013 & 0.009 & 0.003 & 0.000 \\ 
	Deterministic & \textbf{0.020} & 0.010 & 0.003 & 0.000 \\ 
	Stochastic & 0.016 & \textbf{0.018} & \textbf{0.004} & \textbf{0.001} \\ 
	\bottomrule
	\end{tabular}
\end{table}


Let us focus on two interesting issues.
First, most errors for Identity are due to two particular internal files where for one fuzzing setup and for the other fuzzing runs fail (almost) always. 
Deeper analysis revealed that the first failure is caused by specific type conversion issues in combination with annotations that the LLM may insert or not. 
The second failure stems from the fact that the LLM sometimes translates function-like macros to Rust functions, which the fuzzer then cannot find in the C~code.
Second, we see an increase in LLM API errors for stochastic perturbations. This is particularly due to several runs where the model rejects the translation with a BadRequestError, as a response triggered by a perturbation is filtered due to OpenAI's content management policy.

What does \textbf{not} change across all perturbations is the mean number of iterations across all experiments. For each set of Identity, deterministic, and stochastic perturbation experiments, the mean number of iterations is $2.1$.

\subsection{Threats to Validity}

A threat to internal validity of the performance of the translation system is the impact of errors. This was discussed in detail in Sec.~\ref{sec:detailederror}.
The random factor in LLM output may affect the results, however we accounted for this by running our experiments multiple times across various LLMs and code perturbations, and all experiments show consistent results.
Concerning external validity, the experiments were run on only 50~C~files. However, (\textit{i})~we focus on files that capture our target domain (embedded source code) and (\textit{ii})~we sampled from different public and internal sources to increase coverage. Finally, the concrete results on the translation task may not generalize to other LLM-based tasks. They may not even generalize to translation tasks between other language pairs. However, the results do give an impression about the degree of influence of the different factors on LLM results when applied to code.



\section{Conclusion}
\label{sec:conclusion}

This paper evaluated the impact of different factors on translating code from C to Rust via an LLM-based code translation system.
The investigated factors were the influence of feedback loops and repetitions, LLM selection, and code perturbations, and the measured effect was translation performance in terms of compilability and equivalent behavior.
The LLM-based code translation system employs the \textit{generate-and-check} pattern: Generated Rust code is validated through compilation and linting checks and behavioral equivalence testing using differential fuzzing. 

We found that feedback loops and repetitions both significantly enhance translation success, for both non-reasoning and reasoning models.
The first iteration of the feedback loop typically has a higher impact, up to 24\% improvement, than further iterations.
Different models yield varying performance: Among the non-reasoning models, \gptnew\ performed best on the first attempt, while reasoning models have a higher initial performance at higher cost.
However, among non-reasoning models, feedback and repetitions boost \msphi\ from last place for pass@$1$ of the evaluated models, to first place for pass@$5$.
Generally, the differences between models are reduced by the translation system with feedback loop and repetitions.

We also explored the robustness of the translation system against code perturbations, revealing that most perturbations have a small impact on performance yet lead to higher diversity of the results.
Diversity is helpful in \textit{generate-and-check} systems as it can boost the performance of systems when using feedback loops and repetitions.
While we focused on language translation, we assume that this pattern can be leveraged for many LLM-based software engineering use cases.
The results give insights for the development of LLM-based systems with respect to the benefits of choice of LLM and code perturbation robustness.

\bibliographystyle{IEEEtran}
\bibliography{references}

\end{document}